\documentclass[journal]{IEEEtran}
\usepackage{cite}

\ifCLASSINFOpdf
   \usepackage[pdftex]{graphicx}
   \graphicspath{{./images/}}
\else
\fi
\usepackage{tikz}
\usepackage{amsmath}
\usepackage{amsfonts}
\usepackage[%
ruled, 
vlined, 
noend, 
noalgohanging,
]{algorithm2e}

\newcommand{\Comment}[1]{\textit{\scriptsize{\# #1}}\;}
\newcommand{\Input}[1]{\hspace{-10pt}\textbf{Input: }#1\;}
\newcommand{\Output}[1]{\hspace{-10pt}\textbf{Output: }#1\;}
\newcommand{\Blank}{\vspace{3pt}}

\SetNlSty{scriptsize}{}{:}
\SetAlgoNlRelativeSize{0}
\SetNlSkip{3pt}
\SetKwFor{For}{\nl for}{:}{end for}
\SetKwFor{ForAll}{\nl for all}{:}{end for}
\SetKwIF{If}{ElseIf}{Else}{\nl if}{:}{\nl else if}{\nl else}{endif}
\SetArgSty{normal}

\DontPrintSemicolon

\usepackage{tabularx}
\newcolumntype{C}{>{\centering\arraybackslash}X}

\usepackage{pifont}
\newcommand{\cmark}{\textcolor[rgb]{0.4,0.8,0}{\ding{51}}}%
\newcommand{\xmark}{\textcolor{red}{\ding{55}}}%

\RequirePackage{eurosym}
\RequirePackage{siunitx}
\DeclareSIUnit{\eur}{\mbox{\euro}}
\DeclareSIUnit{\ct}{ct}
\DeclareSIUnit{\usdollar}{\small{US}\$}

\newcommand{\tikzin}[1]{%
	\includegraphics{#1.pdf}
}

\usepackage{subcaption}

\begin{document}

\title{Integrating Battery Aging in the Optimization for Bidirectional Charging of Electric Vehicles}

\author{Karl~Schwenk,~\IEEEmembership{Student Member,~IEEE,}
        Stefan~Meisenbacher,
        Benjamin~Briegel,
        Tim~Harr,
		Veit~Hagenmeyer,~\IEEEmembership{Member,~IEEE,}
		and~Ralf~Mikut.
	\thanks{%
		K.~Schwenk is with Mercedes-Benz~AG, Sindelfingen, Germany, and the Institute for Automation and Applied Informatics, Karlsruhe Institute of Technology, Karlsruhe, Germany 
		(e-mail: karl.schwenk@daimler.com; karl.schwenk@kit.edu).
		B.~Briegel, and T.~Harr are with Mercedes-Benz~AG, Sindelfingen, Germany 
		(e-mail: benjamin.briegel@daimler.com; tim.harr@daimler.com).
		S.~Meisenbacher, V.~Hagenmeyer and R.~Mikut are with the Institute for Automation and Applied Informatics, Karlsruhe Institute of Technology, Karlsruhe, Germany 
		(e-mail: stefan.meisenbacher@kit.edu; veit.hagenmeyer@kit.edu; ralf.mikut@kit.edu).}
	\thanks{
		\copyright~2021 IEEE.
		Personal use of this material is permitted.
		Permission from IEEE must be obtained for all other uses, in any current or future media, including reprinting/republishing this material for advertising or promotional purposes, creating new collective works, for resale or redistribution to servers or lists, or reuse of any copyrighted component of this work in other works.}
	}

%

%

%
%
%
\maketitle
\begin{abstract}
Smart charging of Electric Vehicles (EVs) reduces operating costs, allows more sustainable battery usage, and promotes the rise of electric mobility.
%
In addition, bidirectional charging and improved connectivity enables efficient power grid support.
%
Today, however, uncoordinated charging, e.g. governed by users' habits, is still the norm.
Thus, the impact of upcoming smart charging applications is mostly unexplored.
%
We aim to estimate the expenses inherent with smart charging, e.g. battery aging costs, and give suggestions for further research.
%
Using typical on-board sensor data we concisely model and validate an EV battery.
We then integrate the battery model into a realistic smart charging use case and compare it with measurements of real EV charging. 
%
The results show that
i) the temperature dependence of battery aging requires precise thermal models for charging power greater than 7 kW,
ii) disregarding battery aging underestimates EVs' operating costs by approx. 30\%, and
iii) the profitability of Vehicle-to-Grid (V2G) services based on bidirectional power flow, e.g. \textit{energy arbitrage}, depends on battery aging costs and the electricity price spread.
\end{abstract}
\begin{IEEEkeywords}
Electric Vehicle Charging, 
Artificial Neural Network (ANN), 
Vehicle-to-Grid,
Optimization,
Smart Charging,
Electric Vehicles,
Energy Arbitrage
\end{IEEEkeywords}
%
%
%

\ifCLASSOPTIONpeerreview
\begin{center} 
	\textbf{Corresponding Author}
	
\begin{tabular}{l l}
	\hfill & 					\hfill \\
	\textbf{Name}&				Karl Schwenk\\
	\textbf{Title} &			M.Sc. mechanical engineering\\
	\textbf{Postal Address} &	Boehmerwaldstrasse 20, 71263 Weil der Stadt, Germany\\
	\textbf{Phone Number} &		+49 (0) 157 / 71576014\\
	\textbf{E-Mail Address} &	karl.schwenk@kit.edu\\	
	\hfill & 					\hfill \\
\end{tabular}

	\textbf{The present work has not been submitted, accepted or published anywhere prior to this submission.}
\end{center}

\fi
%
\IEEEpeerreviewmaketitle

%
%

\begin{table*}[h]
	\centering
	\caption{Comparison of the proposed work with related works according to distinct technical features, see Sec.~\ref{sec:related_work}.}
	\begin{footnotesize}
		\renewcommand{\arraystretch}{1.2}
		\begin{tabularx}{\textwidth}{l||C|C|C|C|C}
			\hline
			References & 
			A) EV User Perspective Adequately Represented & 
			B) Battery Aging Considered as Monetary Costs & 
			C) Optimization-based Smart Charging Application & 
			D) Thermal Battery Model based on EV Sensor Data & 
			E) Real-world Data for Modeling and Validation \\
			\hline
			\hline
			\cite{Crozier2020, Li2020, Yang2020}																& \xmark & \xmark    & \cmark   & \xmark    & \cmark   \\			
			\cite{Bishop2016, Steffen2020} 																		& \cmark & \cmark   & \xmark    & \xmark    & \xmark    \\
			\cite{Trippe2014,Hesse2019,Tayarani2019,Wan2019,Ebrahimi2020,Gree2020,Vermeer2020,Zhou2020a} 		& \cmark & \cmark   & \cmark   & \xmark    & \xmark    \\
			\cite{Lunz2011, Brinkel2020, Das2020} 																& \cmark & \cmark   & \cmark   & \xmark    & \cmark   \\
			\cite{Petit2016} 																					& \xmark & \xmark    & \xmark    & \cmark   & \cmark   \\
			\hline
			proposed work 																						& \cmark & \cmark   & \cmark   & \cmark   & \cmark   \\
			\hline
		\end{tabularx}
		\label{tab:literature_review}
	\end{footnotesize}
	%
\end{table*}%
\section{Introduction}\label{sec:intro} 
\IEEEPARstart{U}{ndoubtedly}, Electric Vehicles (EVs) are on the rise. 
In this context, Smart Charging (SC), i.e.\ the controlled and coordinated charging of EVs, helps to minimize EV operating cost \cite{Schuller2014} and prolong EV battery life \cite{Li2020}.
To further diminish negative power grid impacts due to EV charging, Vehicle-to-Grid (V2G) that usually leverages bidirectional power flow, becomes important \cite{Saldana2019}.%
\footnote{%
	Note that in this work we consider V2G as a derivative of SC.
}

%
%
%
Diffusion of SC concepts, however, strongly depends on EV user acceptance \cite{Will2016}.
%
Nowadays, EV users typically proceed manual charging in which the battery is fully charged at maximum available power after plugging in. 
For SC, EV users demand similar transparency in terms of operating cost \cite{DAI_EV_REP_2019}. 
%
SC applications usually consist of multi-dimensional decision problems with various objectives.
Instead of simple heuristics whose behavior and effects are easily comprehensible for the EV user, these problems require model-based optimization methods for solving \cite{Tan2016, Das2020}.
%
For this purpose, suitable EV battery models must be developed and validated with real-world data \cite{Das2014}.
Thus, EV operating cost can be correctly determined and the real-world impact of SC can be explored.

%
Aiming to fulfill the aforementioned requirements, we first create a battery model that can be used in production EVs.
The battery model is then integrated into a realistic SC scheme.
Based on validation with real-world data, we draw conclusions for future work on SC.
%
%

%
%
This paper is structured as follows:
Section \ref{sec:related_work} reviews and discusses related work.
All models and their connections are outlined in Section \ref{sec:methods}.
Section \ref{sec:case_study} describes an exemplary SC use case.
Based on this, Section \ref{sec:optimization_results} presents the validation of the single model components and optimization results.
In Section \ref{sec:conclusion} we summarize the major findings and give an outlook on future work.

\section{Related Work} \label{sec:related_work}
To properly define the scope of the proposed work, we proceeded a literature review yielding 30 relevant references.
We classified those based on the following five features (see also Tab.~\ref{tab:literature_review}):
\begin{enumerate}
	
	\item 	
	Diffusion of SC applications strongly depends on EV user acceptance.
	This requires adequate representation of EV operating cost, and/or inclusion of user comfort \cite{Will2016}.
	
	\item 	
	Battery aging is a crucial factor for SC economics that may not be neglected
	\cite{Trippe2014, Hesse2019, Tayarani2019, Nazari2020}.
	In addition, EV users desire transparent EV operating cost, including costs inherent in battery aging \cite{DAI_EV_REP_2019}.
	
	\item 	
	SC applications usually consist of multi-dimensional and multi-objective decision problems that require advanced techniques for solving, 
	e.g.\ model-based optimization \cite{Tan2016, Das2020} or model-free reinforcement learning \cite{Wan2019}.
	
	\item	
	The battery temperature is an important factor for efficiency and battery aging \cite{Thompson2018, Saldana2019}.
	Therefore, a thermal battery model suitable for application with typical EVs sensor data is required.
	
	\item 	
	The complexity of SC use cases, e.g.\ several stakeholders and financial value streams, 
	requires real-world data for both modeling and validation \cite{Das2014}.	
	
\end{enumerate}
Out of the 30 references we further analyzed 
\cite{Crozier2020, Li2020, Yang2020, Bishop2016, Steffen2020, Trippe2014,Hesse2019,Tayarani2019,Wan2019,Ebrahimi2020,Gree2020,Vermeer2020,Zhou2020a, Lunz2011, Brinkel2020, Das2020, Petit2016} 
possessing at least two of the aforementioned features;
a representative excerpt is given in the following.

%
%
Already in 2011 \textit{Lunz et al.}~\cite{Lunz2011} applied a genetic optimization algorithm to bidirectional charging of plug-in hybrid EVs.
Considering dynamic energy tariffs and battery aging costs, EVs' operating cost was reduced compared with uncoordinated charging. 
However, a detailed battery model, e.g. including a thermal model, was not implemented. 

\textit{Brinkel et al.}~\cite{Brinkel2020} proceeded a study on grid reinforcement with respect to limits of low voltage grid transformers and $\text{CO}_{2}$ emissions.
The cost for EV charging was reduced by 13.2\%.
For the study, however, several EVs were aggregated and a simplified battery model was implemented. 

\textit{Das et al.}~\cite{Das2020} set up an optimization-based SC scheme in a micro grid.
The authors aimed to minimize energy costs, battery aging, and $\text{CO}_{2}$ emissions, while maximizing grid utilization.		
To adequately combine the perspectives of all stakeholders, the authors concluded that a multi-objective decision process is required.
Although a 28.1\% reduction of battery aging was reported for some cases, the applied battery model neither considered calendar aging nor the battery's thermal behavior.

\textit{Li et al.}~\cite{Li2020} used particle swarm optimization to design a SC scheme targeting minimal battery aging and grid load fluctuations.
To quantify battery aging, a novel rain-flow cycle counting algorithm was used. 
The approach, however, inadequately represents the EV user perspective, as both variable electricity tariffs and monetary battery aging costs were neglected.

In \cite{Petit2016}, \textit{Petit et al.} set up a SC application using an empirical battery aging model that considers electro-thermal influences.
The case study for validation, however, was based on a heuristic that neglects important factors from an EV user perspective, e.g.\ energy and battery aging costs.

To the best of the authors' knowledge, none of the related works found in the literature considers all five features as indicated in Tab.~\ref{tab:literature_review}.
%
Hence, we summarize the contributions of this paper as follows:
\begin{itemize}
	
	\item 	We use data of real-world charging events to design and validate a vehicle-specific battery model;
	this comprises the battery's electrical, thermal, and aging behavior. 
	All models can be operated with inputs from typical onboard sensors in production EVs.
	
	\item  	This battery model is integrated in an optimization-based SC scheme.
	We then use Discrete Dynamic Programming (DDP) as a robust method for solving.
	For validation, we consult data from real-world charging events and historical electricity market prices.
	
	\item 	To support future work on SC, we derive application-dependent suggestions on
	i) the necessity of thermal battery models, 
	ii) the significance of battery aging costs, and
	iii) suitable electricity tariffs for profitable V2G applications. 
	
\end{itemize}
\section{Models of a Smart Charging Scheme}\label{sec:methods}
\begin{figure*}[!t]
	\begin{scriptsize}
		\centering
		\begin{subfigure}{0.45\textwidth}
			\centering
			\tikzin{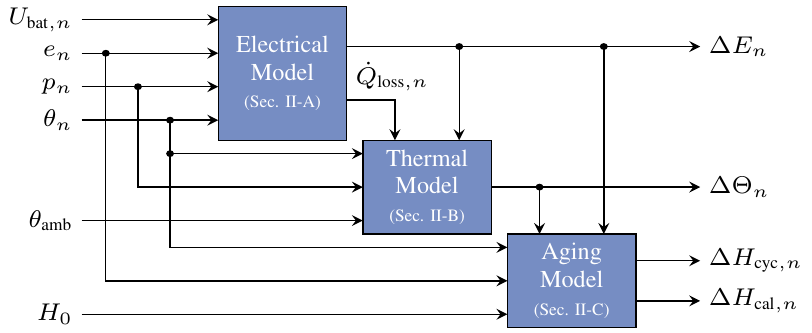}
			\caption{Battery model and its components}		
			\label{sfig:battery_model}
		\end{subfigure}
		\hspace{0.04\textwidth}
		\begin{subfigure}{0.45\textwidth}
			\centering
			\tikzin{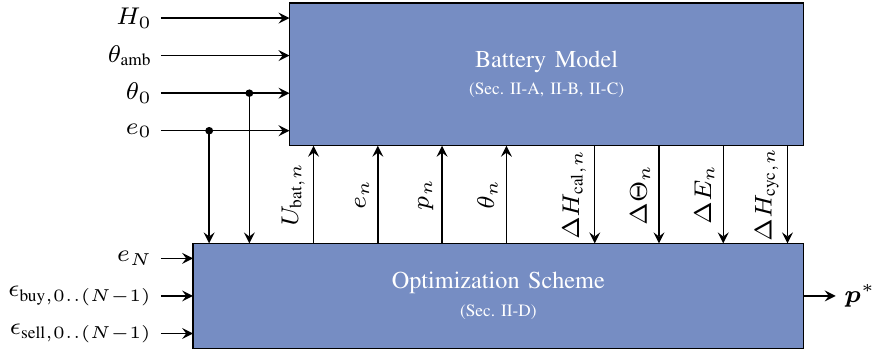}
			\caption{Optimization scheme for smart charging}
			\label{sfig:optimization_scheme}		
		\end{subfigure}
		\hspace{0.04\textwidth}
		\caption{
			Layout of used models for smart charging application;
			battery model (Sec.~\ref{ssec:electrical_model}, \ref{ssec:thermal_model}, \ref{ssec:aging_model}) (a);
			optimization scheme (Sec.~\ref{ssec:optimization_scheme}) (b).
		}
		\label{fig:model_layout}
		\hrulefill
	\end{scriptsize}
\end{figure*}
First, we introduce the notation
for a charging event starting at arrival time $t_0$ and ending at departure time $t_N$.
%
%
%
%
%
%
The time horizon $\left[t_0, t_N\right]$ is divided into $N$ time intervals of duration $\Delta t$ and $N+1$ states.
Accordingly, we define the set of intervals 
\begin{equation}
	\mathcal{N} = [0, N-1] \subset \mathbb{N}.
\end{equation}
Each time interval $n \in \mathcal{N}$ starts at time $t_{n}$ and ends at time $t_{n+1}$.
Each battery state at $t_n \in \left[t_0, t_N\right]$ is characterized by the battery energy $e_n$, normalized as State of Charge (SOC), and the battery temperature $\theta_n$.
The charging power $p_n$ is assumed to remain constant throughout a single time interval $n, \forall n \in \mathcal{N}$.%
\footnote{
	Note that $p_n$ represents the gross charging power consumed from the charging station without conversion losses.
}

To represent the battery's charging behavior, we implement an electrical, thermal, and aging model, see Fig.~\ref{sfig:battery_model}.%
\footnote{%
	These models can be vehicle-, vehicle-type-, or battery-specific, thus limiting a generic reuse.
	General initialization followed by incremental adaption, however, is conceivable. 
}
We refer to the combination of these three models as the battery model.
The perspective use of this battery model in production EVs limits the model inputs to typical onboard sensor data.
Further, an optimization scheme as shown in Fig.~\ref{sfig:optimization_scheme} serves to calculate an optimal charging power trajectory $\boldsymbol{p}^{*}$ for a single charging event.

\subsection{Electrical Model} \label{ssec:electrical_model}
The energy level of the EV battery---and thus the SOC---changes with surrounding influences, especially the charging power $p_n$. 
An electrical battery model, see Fig.~\ref{sfig:battery_model}, helps to calculate the battery energy evolution.
For this purpose, the energy throughput of the battery
\begin{equation}
	\Delta E_n = e_{n+1} - e_{n}, \forall n \in \mathcal{N},
\end{equation}
is estimated for a given 
time interval $n$, 
battery temperature $\theta_{n}$, 
battery energy $e_{n}$, 
battery's terminal voltage $U_{\text{bat}, n}$, and 
charging power $p_n$.

\begin{figure}[!t]
	\centering
	\tikzin{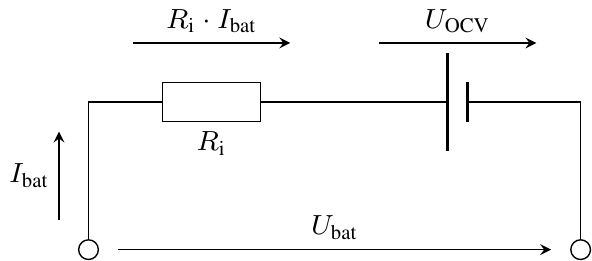}
	\caption{%
		Equivalent circuit model of an EV battery for low-dynamic operation with internal resistance $R_{\text{i}}$ and voltage source $U_{\text{OCV}}$.
	}
	\label{fig:equivalent_circuit}
\end{figure}

We abstract the EV battery with an Equivalent Circuit Model (ECM) consisting of a voltage source $U_\text{OCV}$ serially connected with the internal resistance $R_\text{i}$, see Fig.~\ref{fig:equivalent_circuit}.%
\footnote{%
	This model represents dedicated power electronics, hence it is vehicle-specific.  	
}
Due to the low dynamics of EV charging, more complex models, e.g. resistor-capacitor-pairs or electro-chemical models, are not required \cite{Plett2004}.

Both $U_{\text{OCV}, n}$ and $R_{\text{i}, n}$ depend on the battery temperature $\theta_{n}$ and the battery energy $e_{n}$;
we assume their values to be constant throughout a single time interval $n$.
We obtain $R_{\text{i}}$ from a look-up table and use the measured terminal voltage $U_{\text{bat}, n}$ to obtain the open-circuit voltage of the battery
\begin{equation}
	U_{\text{OCV}, n} = U_{\text{bat}, n} - R_{\text{i}, n} \cdot I_{\text{bat}, n}.
	\label{eq:u_ocv}
\end{equation}

With the battery current $I_{\text{bat}} > 0$ during charging, $U_\text{bat} > U_\text{OCV}$ for charging and $U_\text{bat} < U_\text{OCV}$ for discharging. 
Substituting the terminal voltage with 
\begin{equation}
	U_{\text{bat}, n} = \frac{p_n}{I_{\text{bat}, n}},
\end{equation}
and solving \eqref{eq:u_ocv}, we obtain the battery current%
\footnote{%
	Despite two possible solutions only the greater one is physically feasible. 
}
\begin{equation}
	I_{\text{bat}, n}= \frac{-U_{\text{OCV}, n} + \sqrt{U_{\text{OCV}, n}^{2} + 4R_{\text{i}, n} \cdot p_n}}{2R_{\text{i}, n}}.
\end{equation}
The \textsc{Ohm}ic \cite{Millikan1917} losses within the battery amount to 
\begin{equation}
	\dot{Q}_{\text{loss},n} = R_{\text{i}, n} \cdot I_{\text{bat}, n}^2.
	\label{eq:q_loss}
\end{equation}
Given the charging power $p_n$, we obtain the energy throughput 
\begin{equation}
	\Delta E_n = \Delta t \cdot \left(p_n - \dot{Q}_{\text{loss},n}\right).
	\label{eq:energy_transition_ecm}
\end{equation}
Note that $\dot{Q}_{\text{loss},n}>0$ occurs both while charging and discharging.
Hence, it decreases $|\Delta E_n|$ during charging and increases $|\Delta E_n|$ during discharging.

\subsection{Thermal Model} \label{ssec:thermal_model}
Both the internal battery parameters and battery aging depend on the battery temperature.
Therefore, a thermal battery model, as shown in Fig.~\ref{sfig:battery_model}, estimates the change in battery temperature 
\begin{equation}
	\Delta \Theta_n = \theta_{n+1} - \theta_{n},
\end{equation}
for a given 
time interval $n$,
battery temperature $\theta_{n}$, 
battery energy $e_{n}$, and
charging power $p_n$.%
\footnote{%
	Due to heat exchange with surrounding components, thermal battery models are mostly vehicle-type-specific.	
}
As the thermal behavior of EV batteries follows complex, non-linear processes, e.g. electro-chemical heat sources or sinks, we use a data-driven modeling approach.

First, we explore time series data from real charging events of batteries installed and operated in EVs;
we discretize this data with $\Delta t = 5 \text{ min}$ to obtain single training samples for each time step. 
Based on the \textsc{Spearman} \cite{Spearman1904} correlation coefficient, we screen out irrelevant features. 
We then perform mean and variance normalization to ensure a proper model training.
The available training data underrepresents discharging, i.e. $p_n < 0$.
However, as experiments with the installed battery show similar $R_{\text{i},n}$ for charging and discharging, we consider it acceptable to use absolute values for $p_n$ and $I_{\text{bat},n}$.

For the machine learning models, we compare a linear regression model and different Artificial Neural Network (ANN) models 
(multi layer perceptron with sigmoid activation function, learning rate of 0.001).
As the optimal ANN model architecture 
(number of hidden layers, number of neurons per hidden layer) 
may vary for different input features, we use \textit{grid-search}~\cite{ScikitLearn2011} to obtain the best performing model architecture. 
Applying a five-fold cross validation, we select the features 
\begin{equation}
	\lbrace p_n, \dot{Q}_{\text{loss}, n}, \Delta E_n, \theta_{n} \rbrace,
	\label{eq:features_thermal_model}
\end{equation}
to estimate $\Delta \Theta_n$.
Note that we engineer the additional feature $\dot{Q}_{\text{loss}, n}$ calculated in \eqref{eq:q_loss} based on domain knowledge.
Hence, we uncover hidden relations for the machine learning algorithm (\textit{gray-box} approach).
We implement all models in \textit{Python} \cite{Python2019} using \textit{SciKit-Learn}~\cite{ScikitLearn2011} for linear regression and \textit{Keras}~\cite{Keras2015} for ANNs. 

\subsection{Battery Aging Model}
\label{ssec:aging_model}
Irreversible physical and electro-chemical degradation processes (\textit{battery aging}) cause the EV's usable driving range and monetary value to decrease.
To quantify battery aging, the State of Health (SOH)
\begin{equation}
	H = \frac{e_{\text{max}}}{e_{\text{nom}}} \leq 1,
\end{equation}
indicates the maximum available storage capacity $e_{\text{max}}$ compared with the nominal storage capacity $e_{\text{nom}}$.
Calculating the evolution of $e_{\text{max}}$ requires a battery aging model as described in the following.%
\footnote{
	Note that this model represents battery-specific aging characteristics that may differ for different types of battery cells;
	the proposed framework allows to accordingly replace the aging model, e.g. with models as in \cite{Xu2018}.
}

%
Charging and discharging causes the battery's anode and cathode to decay (\textit{cyclic aging}).
Among other processes, a loss of active lithium material occurs due to mechanical stress, see also \cite{Woody2020};
For the battery cells used in this study, the cyclic aging increment
\begin{equation}
	\Delta H_{\text{cyc},n} = \beta_{\text{A}} \cdot \left|\Delta E_{n}\right|^{\beta_{\text{B}}},
	\label{eq:cyclic_aging}
\end{equation}
only depends on the absolute energy throughput $\Delta E_{n}$.

%
Additionally, high battery temperature and SOC cause degradation of both active and inactive battery components, see also \cite{Woody2020}.
Hence, the battery capacity fades over time, regardless of the energy throughput (\textit{calendar aging}).
For the battery cells used in this work, we describe the calendar aging increment
%
%
\begin{align}
	\label{eq:calendar_aging}
	& \Delta H_{\text{cal},n} = \\
	& 1 - H_0 + \beta_{\text{C}} \exp\left(\frac{\beta_{\text{D}}}{273 + \theta_{n}} + \beta_{\text{E}} e_{n} \right) \cdot (\Delta t + \tau_n)^{\beta_{\text{F}}},\nonumber
\end{align}
based on \textsc{Arrhenius} \cite{Gold2019} curves.
Here, $H_0$ is the SOH at the beginning of the charging event.%
\footnote{
	$H_0$ serves as a reference for all time steps of one charging event, as calendar aging occurs on larger time scales (years) than charging (hours).
}
Furthermore,
\begin{equation}
	\tau_n = \left( \frac{H_0 - 1}{\beta_{\text{C}} \exp\left(\frac{\beta_{\text{D}}}{273 + \theta_{n}} + \beta_{\text{E}} e_{n} \right)} \right)^{1/{\beta_{\text{F}}}},
\end{equation}
represents the equivalent battery age for each time interval $n \in \mathcal{N}$ as a function of $H_0$.

%
Both model characteristics and parameters are estimated from extensive cell tests at varying conditions, e.g.\ battery energy and temperature.
Hence, the detailed parameters $\beta_{\text{A..F}}$ depend on battery (cell) type and are confidential. 
Prospectively, also an implicit representation, e.g.\ via machine learning approaches as in \cite{Berecibar2019} is conceivable.

\subsection{Optimization Scheme}\label{ssec:optimization_scheme}
Calculating a charging event's power trajectory requires appropriate algorithms, e.g.\ optimization-based SC scheduling schemes.  
We therefore modify the vehicle- and battery-independent optimization scheme from \cite{Das2020} such that
\begin{subequations}
	\begin{align}
		\label{eq:objective}
		&\min\limits_{\boldsymbol{p} \in \mathbb{R}^N} \sum_{\forall n \in \mathcal{N}} J_{\text{E}, n}(p_n, \epsilon_{\text{buy}, n}, \epsilon_{\text{sell}, n}) + J_{\text{D}, n}(\theta_n, e_n, H_0) \\
		& \nonumber \\
		&\text{subject to} \nonumber \\
		\label{eq:decision_variable}
		&\underline{\boldsymbol{p}}\leq \boldsymbol{p} \leq \overline{\boldsymbol{p}}, \quad \boldsymbol{p} \in \mathbb{R}^N, &\\
		\label{eq:battery_energy}
		&\underline{\boldsymbol{e}}\leq \boldsymbol{e} \leq \overline{\boldsymbol{e}}, \quad \boldsymbol{e} \in \mathbb{R}^{N+1}, \\
		\label{eq:constraint_initial_energy}
		&\underline{e}_{0} = e_0 = \overline{e}_{0},\\
		\label{eq:constraint_target_energy}
		&\underline{e}_{N} = e_N = \overline{e}_{N},\\
		\label{eq:battery_temperature}
		&\underline{\boldsymbol{\theta}}\leq \boldsymbol{\theta} \leq \overline{\boldsymbol{\theta}}, \quad \boldsymbol{\theta} \in \mathbb{R}^{N+1},\\
		\label{eq:constraint_initial_temperature}
		&\underline{\theta}_{0} = \theta_0 = \overline{\theta}_{0},\\
		\label{eq:energy_transition}	
		&e_{n+1} = e_{n} + \Delta E \left(e_{n}, \theta_{n}, p_n \right), \quad \forall n \in \mathcal{N},\\
		\label{eq:temperature_transition}	
		&{\theta}_{n+1} = {\theta}_{n} + \Delta \Theta\left(e_{n}, p_n\right), \quad \forall n \in \mathcal{N}.
	\end{align}
	\label{subeq:optimization_problem}
\end{subequations}
Figure~\ref{sfig:optimization_scheme} shows the interaction of the optimization scheme and the proposed battery model.
The components \eqref{eq:objective}-\eqref{eq:temperature_transition} and the solving method are described in the following.

\subsubsection{Cost Functions}
The optimization objective \eqref{eq:objective} is to minimize the sum of energy costs $J_{\text{E}, n}$ and aging costs $J_{\text{D}, n}$ over all time intervals $n \in \mathcal{N}$.

%
%
%
To consider the costs for charging electric energy, we define the energy cost function
\begin{equation}
	J_{\text{E}, n} = \left\{
	\begin{array}{lr}
		J_{\text{E}, n}^{+}, & \forall p_n \geq 0, \\ 
		J_{\text{E}, n}^{-}, & \forall p_n < 0,
	\end{array}
	\right.
	\label{eq:energy_cost}
\end{equation}
with the energy expenses
$J_{\text{E}, n}^{+} = p_n \cdot \Delta t \cdot \epsilon_{\text{buy}, n}$,
and the energy rewards
$J_{\text{E}, n}^{-} = p_n \cdot \Delta t \cdot \epsilon_{\text{sell}, n}$. 
%
%
%
For $p_n \geq 0$, the EV battery is charged at the electricity price $\epsilon_{\text{buy}, n}$.
With $p_n < 0$, $\epsilon_{\text{sell}, n}$ corresponds to the price of selling energy back to the grid.
Both $\epsilon_{\text{buy}, n}$ and $\epsilon_{\text{sell}, n}$ are given and assumed to be deterministic.

%
%
%
Battery aging also contributes to the total operating cost, as the EV's monetary value depends on the maximum usable battery capacity $e_{\text{max}}$.
Based on $\Delta H_{\text{cyc},n}$ and $\Delta H_{\text{cal},n}$ (Sec.~\ref{ssec:aging_model}) we define the battery aging costs
\begin{equation}
	J_{\text{D}, n} = \underbrace{\Delta H_{\text{cyc},n} \cdot \frac{V_{\text{EV}}}{H_{\text{EV}}}}_{J_{\text{D}, n}^{\text{cyc}}} 
	+ \underbrace{\Delta H_{\text{cal},n} \cdot \frac{V_{\text{EV}}}{H_{\text{EV}}}}_{J_{\text{D}, n}^{\text{cal}}},
	\label{eq:aging_cost}
\end{equation}
with the cyclic aging costs $J_{\text{D}, n}^{\text{cyc}}$ and the calendar aging costs $J_{\text{D}, n}^{\text{cal}}$.%
\footnote{%
	Given the non-linearity in \eqref{eq:cyclic_aging} and \eqref{eq:calendar_aging}, \eqref{eq:aging_cost} also introduces non-linearity to \eqref{subeq:optimization_problem}.
}
Here, $V_{\text{EV}}$ denotes the battery value loss due to the capacity loss $H_{\text{EV}}$ during the battery's automotive application (\textit{first life}).
In particular, $V_{\text{EV}}$ is the difference of the battery's production price and residual value in a \textit{second life} market.
Note that \eqref{eq:aging_cost} only accounts for aging caused throughout the charging event.
To include battery aging for trips in between charging events in future work, a superordinate scheme as in \cite{Schwenk2019} could determine the optimal target energy $e_N$.

\subsubsection{Decision Variable}
To obtain the optimal charging power trajectory, we define the decision variable
\begin{equation}
	\boldsymbol{p} = (p_0, p_1, ..., p_{N-1})^{\top}\in \mathbb{R}^{N},
\end{equation}
with the charging power $p_n$ in all time intervals $n \in \mathcal{N}$.
Evaluating \eqref{eq:decision_variable} component-wise represents the power limitations with the upper bounds $\overline{\boldsymbol{p}}$ and lower bounds $\underline{\boldsymbol{p}}$.
We assume these bounds to be constant throughout a single charging event.
Both $\overline{\boldsymbol{p}}$ and $\underline{\boldsymbol{p}}$ are known at the time of computation and predefined e.g. by grid load constraints, charging stations, or EV power electronics.

\subsubsection{State Variables}
To compute the SOC of the battery, we define the state variable
\begin{equation}
	\boldsymbol{e} = (e_0, e_1, e_2, ..., e_N)^{\top}\in \mathbb{R}^{N+1},
\end{equation}
representing the battery energy trajectory throughout a charging event.
Reading \eqref{eq:battery_energy} component-wise reveals the energy limitations $\overline{\boldsymbol{e}}$ and $\underline{\boldsymbol{e}}$ that are constant and known at the time of computation.
Their values are determined by physical restrictions, i.e.\ battery capacity, and/or the EV user preferences, e.g.\ a minimum SOC.%
\footnote{%
	Note that the use of an electrical model (Sec.~\ref{ssec:electrical_model}) in \eqref{eq:energy_transition} implicitly respects the battery's voltage constraints.
}
We specify $e_0$ as the battery energy at the beginning of the charging event in \eqref{eq:constraint_initial_energy}.
The desired battery energy at departure $e_N$ is determined by the user in \eqref{eq:constraint_target_energy};
prospectively, also a previous calculation of $e_N$ as in \cite{Schwenk2019} can be used.

For the battery temperature we similarly define the state variable
\begin{equation}
	\boldsymbol{\theta} = (\theta_0, \theta_1, \theta_2, ..., \theta_N)^{\top}\in \mathbb{R}^{N+1}.
\end{equation}
The temperature limits $\overline{\boldsymbol{\theta}}$ and $\underline{\boldsymbol{\theta}}$ in \eqref{eq:battery_temperature} restrict $\boldsymbol{\theta}$ to be within a safe operating window given by the battery management system. 
Both $\overline{\boldsymbol{\theta}}$ and $\underline{\boldsymbol{\theta}}$ are constant and known at the time of computation.
Additionally, the battery temperature $\theta_0$ at the beginning of the charging event is given in \eqref{eq:constraint_initial_temperature}.

\subsubsection{Battery Dynamics}
To represent the evolution of the battery energy $e_n$ we formulate \eqref{eq:energy_transition}.
Particularly, the electrical model in Sec.~\ref{ssec:electrical_model} is used to calculate $e_{n+1}$ for each time interval $n \in \mathcal{N}$ 
(see also Fig.~\ref{sfig:optimization_scheme}).
In a similar manner, \eqref{eq:temperature_transition} describes the battery's thermal behavior based on the thermal model in Sec.~\ref{ssec:thermal_model}.

\subsubsection{Method of Solving} \label{sssec:solving_method}
As the non-linear models in \eqref{eq:energy_transition}-\eqref{eq:temperature_transition} increase the complexity of \eqref{subeq:optimization_problem}, we use DDP \cite{ArtLew2006} for solving;
the detailed solving Algorithms \ref{alg:backward_ind_init_ddp}-\ref{alg:forward_int_ddp} are given in the appendix.
First, Algorithm~\ref{alg:backward_ind_init_ddp} initializes cost grid $\boldsymbol{\mathfrak{J}}$, spanned by $\boldsymbol{e}, \boldsymbol{\theta}$ and time intervals $n \in \mathcal{N}$.
%
Then, Algorithm~\ref{alg:backward_ind_ddp} runs backwards ($n\!=\!N..0, \forall n\!\in\!\mathcal{N}$) to update $\boldsymbol{\mathfrak{J}}$ for all possible $e_i\!\in\!\boldsymbol{e}_{\text{d}}$ and $\theta_j\!\in\!\boldsymbol{\theta}_{\text{d}}$.
Also, all corresponding optimal actions $\boldsymbol{\mathfrak{P}}$ are determined (\textit{backward induction}).
To avoid infeasible trajectories, a penalty value $\lambda$ is assigned to the according value in $\boldsymbol{\mathfrak{J}}$, if violating any constraint \eqref{eq:decision_variable}-\eqref{eq:temperature_transition}.
%
Finally, Algorithm~\ref{alg:forward_int_ddp} integrates forward in time ($n\!=\!0..N, \forall n\!\in\!\mathcal{N}$), starting from the initial values $e_{0}$ and $\theta_{0}$.
For each $n \in \mathcal{N}$, the cost-optimal action is therefore taken from $\boldsymbol{\mathfrak{P}}$ based on the current state $e_{n}$ and $\theta_{n}$.
Thus, the globally optimal charging power trajectory $\boldsymbol{p}^*$ is obtained (\textit{forward integration}) \cite{ArtLew2006}.

\section{Case Study}\label{sec:case_study}
From ten real-world EVs equipped with cloud-connected data loggers, we obtain onboard measured time series data of 279 unidirectional charging events covering a full year \cite{Schwenk2019a};
we discretize this data with $\Delta t = 5 \text{ min}$.
For each interval, we calculate $\Delta E$ and $\Delta \Theta$ with the battery models in Sec.~\ref{ssec:electrical_model} and \ref{ssec:thermal_model}.
We therefore use the mean charging power, and battery temperature and energy at the beginning of each time interval. 
Then, we determine the Root Mean Squared Error (RMSE) of actual and estimated $\Delta E$ and $\Delta \Theta$ for each time interval (\textit{local error}).
Furthermore, we quantify the error propagation when repeatedly applying the battery models 
(i.e.\ estimate $\Delta E$ and $\Delta \Theta$ based on estimations of previous time interval);
we therefore calculate the Mean Absolute Error (MAE) of actual and estimated $e_N$ and $\theta_N$ at the end of each charging event (\textit{global error}).

\begin{figure}[!t]
	\begin{footnotesize}
		\centering
		\tikzin{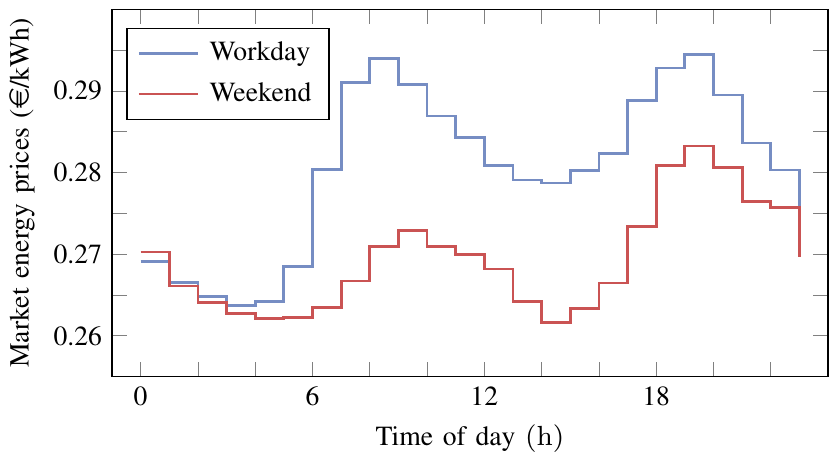}
		\caption{%
			Characteristic price profiles $\epsilon_{\text{buy}} = \epsilon_{\text{sell}}$ of retail electricity for workdays and weekends supplemented by 0.188 \euro/kWh fees and 19\% taxes \cite{EPEX}.%
		}
		\label{fig:energy_price_profile}
	\end{footnotesize}
\end{figure}

To evaluate the optimization scheme (Sec.~\ref{ssec:optimization_scheme}), we select 45 real charging events that have sufficient duration (${t_N \! - \!t_0 \geq \SI{2}{\hour}}$) and set the parameters:
\begin{align*}
	\begin{footnotesize}
		\begin{array}{llll}
			\underline{{p}} = \SI{-50}{\kilo\watt} 	& \underline{{e}} = 8\text{ kWh} 			& \underline{{\theta}} = \SI{-25}{\degreeCelsius} 	&	H_{\text{EV}} = \SI{20}{\percent}\\
			\overline{{p}} = \SI{50}{\kilo\watt}  	& e_{\text{nom}} ={{e}} = 80\text{ kWh} 	& \overline{{\theta}} = \SI{60}{\degreeCelsius} 	& V_{\text{EV}} = \SI{6080}{\eur}\\
		\end{array}
	\end{footnotesize}
\end{align*}
In addition, the conditions of each charging event determine the values of 
$\underline{e}_{0} = \overline{e}_{0}$ (battery energy at arrival), 
$\underline{e}_{N} = \overline{e}_{N}$ (battery energy at departure), and
$\underline{\theta}_{0} = \overline{\theta}_{0}$ (battery temperature at arrival).
After solving \eqref{subeq:optimization_problem} (see Sec.~\ref{sssec:solving_method}), we compare the operating cost in three modes:
\begin{itemize}
	
	\item 	\textbf{Mode I:}
	No optimization, 
	calculate energy and aging costs for measured energy and temperature profile.
	
	\item 	\textbf{Mode II:}
	Optimize for energy costs only, 
	calculate aging costs afterward.
	
	\item 	\textbf{Mode III:} 
	Optimize for both energy and aging costs.
\end{itemize}
We use historic hourly electricity market prices of 2018 to initially set $\epsilon_{\text{buy}, n} = \epsilon_{\text{sell}, n}$. 
To attain a representative price level of private customers, we supplement the market prices with typical fees (0.188 \euro/kWh) and taxes (19\%).
Then, we average the price curves over all workdays and all weekends to level out price peaks occurring in case of electricity over- or underproduction.
We thus obtain two characteristic hourly price tables for $\epsilon_{\text{buy}}$ and $\epsilon_{\text{sell}}$, see Fig.~\ref{fig:energy_price_profile}, to evaluate the average profitability of SC.

Note that the duration of each charging event represents deterministic EV user behavior; 
future work will also consider stochastic influences.
Furthermore, $\boldsymbol{p}^*$ is calculated once at the beginning of each charging event;
prospectively, an ongoing charging event may be adapted to dynamic changes, i.e. departure time, target energy $e_N$, or electricity tariff, via a user interface with a system as in \cite{Meisenbacher2021}.

\begin{figure}[!t]
	\begin{footnotesize}
		\centering
		\tikzin{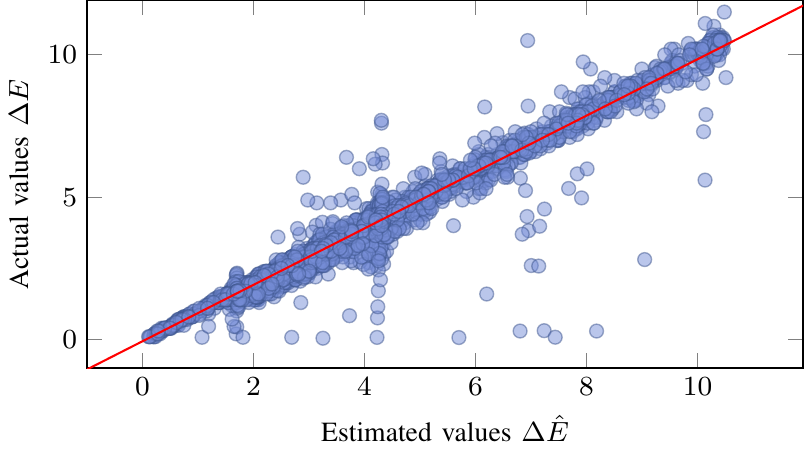}
		\caption{%
			Local error of electrical battery model 
			(Sec.~\ref{ssec:electrical_model}), 
			estimates per time interval $\Delta t = 5 \text{ min}$;
			the red line indicate ideal model behavior.
		}		
		\label{sfig:local_error_ecm}
	\end{footnotesize}
\end{figure}
\begin{figure}[!t]
	\begin{footnotesize}
		\centering
		\tikzin{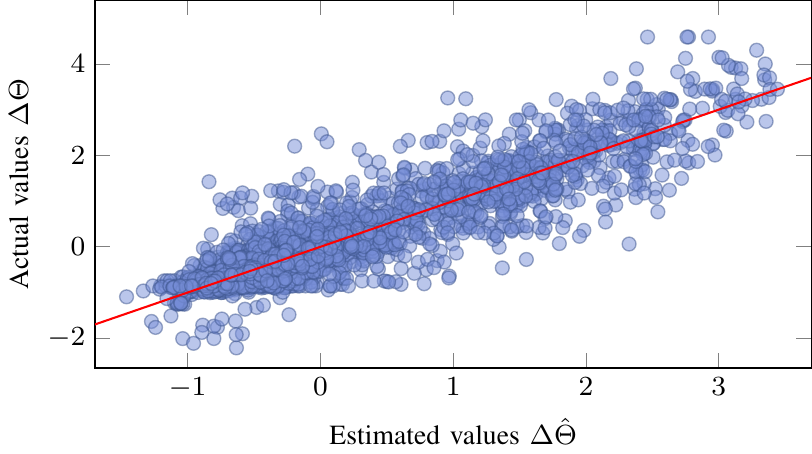}
		\caption{%
			Local error of best-performing data-driven thermal model 
			(ANN, 2 hidden layers, 10 neurons each, Sec.~\ref{ssec:thermal_model}), 
			estimates per time interval $\Delta t = 5 \text{ min}$;
			the red line indicate ideal model behavior.
		}
		\label{sfig:local_error_ann}
	\end{footnotesize}
\end{figure}
%
%
%

%
%
%
%
%
%

%
%
%
\section{Results}\label{sec:optimization_results}
\subsection{Validation of the Battery Model}
Table~\ref{tab:model_errors} presents the validation results of the battery model (Sec.~\ref{ssec:electrical_model}, \ref{ssec:thermal_model}).

\begin{table}[!b]
	\caption{Local and global error of electrical and thermal battery models (Sec.~\ref{ssec:electrical_model}, \ref{ssec:thermal_model}).}
	\label{tab:model_errors}
	\centering
	\renewcommand{\arraystretch}{1.1}
	\begin{tabular}{r||c|c}
		\hline
		\textbf{Model} & 				\textbf{Local Error} & 			\textbf{Global Error}\\
		\hfill &						(RMSE) & 						(MAE)\\
		\hline
		\hline
		Electrical ECM 					& $0.35\text{\% SOC}$ & 		$2.37 \text{\% SOC}$\\	
		Constant Battery Temperature 	& $0.72\text{ K}$ & 			$7.57\text{ K}$\\
		Linear Regression Thermal Model & $0.76\text{ K}$ & 			$4.18\text{ K}$\\
		ANN Thermal Model 				& $0.29\text{ K}$ & 			$1.96\text{ K}$\\
		\hline
	\end{tabular}
\end{table}
\subsubsection{Validation of the Electrical Model}\label{ssec:validation_ecm}
The electrical model (Sec.~\ref{ssec:electrical_model}) yields an RMSE of $0.35\text{\% SOC}$ for single time intervals (local error, see Sec.~\ref{sec:case_study}).
Hence, the model estimations $\Delta \hat{E}$ mostly fit the actual values $\Delta E$ (Fig.~\ref{sfig:local_error_ecm}).
This can be achieved, as both $U_{\text{OCV}, n}$ and $R_{\text{i}, n}$ are chosen from a look-up table for each time interval $n$ depending on $e_n$ and $\theta_n$.
Furthermore, high-dynamic changes of the battery energy are leveled out, as they occur on smaller time scales than the chosen $\Delta t = 5 \text{ min}$ \cite{Plett2004}.
If charging the battery to $e_{\text{max}}$, the battery management system corrects the characteristic SOC curve towards the end of the charging event;
hence, a few outliers occur, e.g. with $\Delta E \cong 0$.

At the end of all charging events, the mean SOC deviation $E_N - \hat{E}_N$ is $2.37 \text{\% SOC}$ (global error), see Tab.~\ref{tab:model_errors};
this equals an acceptable driving range deviation of approx. \SI{7.6}{\kilo\meter}. 
We thus consider the accuracy of the electrical model as sufficient and deem $\Delta t = 5 \text{ min}$ and the ECM to be suitable for our case study.

\subsubsection{Validation of the Thermal Model}\label{ssec:validation_thermal}
To benchmark the thermal battery model (Sec.~\ref{ssec:thermal_model}), we assume constant battery temperature,
i.e.\ $\theta_n\!=\!\theta_{0}, \forall n \!\in \! \mathcal{N}$ and $\Delta \hat{\Theta}_n\!=\!0.0, \forall n\!\in\!\mathcal{N}$.
In comparison with real charging events, assuming constant battery temperature yields an RMSE of $0.72\text{ K}$ for single time intervals (local error, see Sec.~\ref{sec:case_study}) and an MAE of $7.57\text{ K}$ at the end of a charging event (global error), see Tab.~\ref{tab:model_errors}.

As this result emphasizes the need for appropriate modeling, we tested different data-driven thermal models and hyperparameters, see Sec.~\ref{ssec:thermal_model}.
A linear regression model still misrepresents the battery's thermal behavior (see Tab.~\ref{tab:model_errors}), e.g. due to hidden electro-chemical processes. 

More advanced ANN models, however, can capture the apparent non-linearity.%
\footnote{%
	For the sake of brevity we only report the best-performing ANN model (two hidden layers, ten neurons each, sigmoid activation function). 		
}
Using a distinct test data set, the ANN thermal model yields an RMSE of $0.29\text{ K}$ for single time intervals (local error), see Fig.~\ref{sfig:local_error_ann}.
The MAE at the end of a charging event is $1.96\text{ K}$ (global error), see Tab.~\ref{tab:model_errors}.
For our case study we deem this accuracy as sufficient (see also Sec.~\ref{ssec:result_thermal_modeling_effects}).
Yet, the input features of the data-driven thermal model seem to lack further influences on the battery's thermal behavior.
The input features could therefore be enhanced, e.g. by internal cell temperatures or ambient conditions (temperature, sun radiation, wind).
Note that typical EV onboard data, however, does not yet provide this information. 

\subsection{Operating Cost Evaluation}
\begin{figure}[!t]
	\begin{footnotesize}
		\centering
		\tikzin{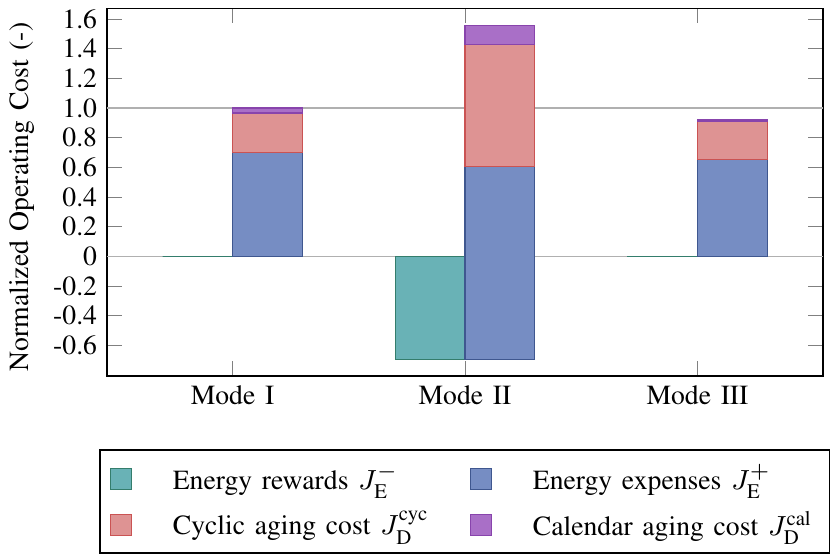}
		\caption{
			Normalized operating cost and its components of 45 real charging events; 
			Mode I (no optimization), 
			Mode II (energy costs optimization),
			Mode III (energy and aging costs optimization).
		}
		\label{fig:operating_cost_comparison}
	\end{footnotesize}
\end{figure}

Figure~\ref{fig:operating_cost_comparison} shows the operating cost components for all three modes described in Sec.~\ref{sec:case_study} normalized against the operating cost of Mode I.
In average, Mode III yields 7.8\% lower operating cost compared with Mode I; 
similar results can be found in the literature, e.g. 5.4\% in \cite{Tayarani2019} and 13.2\% with simplifications in \cite{Brinkel2020}.
Although $\underline{p} < 0$ in \eqref{eq:decision_variable}, i.e. discharging the EV battery is possible, no energy rewards $J_{\text{E}}^{-}$ can be observed in Mode III.
This implies that $J_{\text{E}}^{-}$ does not compensate for round-trip energy losses (charging and discharging) and aging costs.

Disregarding battery aging in average underestimates the total operating cost in Mode I by 30.1\%; 
in \cite{Trippe2014} even up to 52\% is reported.
This becomes apparent when applying Mode II: the optimization scheme utilizes price differences throughout the charging events to generate energy rewards.
Thus, the energy costs \eqref{eq:energy_cost} decreases by 13.3\% compared with Mode I.
Calculating the battery aging costs \eqref{eq:aging_cost} afterward, however, yields 55.8\% higher total operating cost.
Repeatedly charging and discharging the battery increases the battery temperature $\theta$ and causes the calendar aging costs $J_{\text{D}}^{\text{cal}}$ to rise in Mode II.
\textit{Trippe et al.}~\cite{Trippe2014} report an even mode drastic result for this setup: 8\% electricity costs reduction, but a threefold increase of battery aging costs.
Hence, we conclude that especially for charging with the allocation of V2G services--in this case \textit{energy arbitrage}--battery aging must not be neglected.

\subsection{Effects of Thermal Modeling} \label{ssec:result_thermal_modeling_effects}
Including advanced thermal models, e.g. ANNs, increases the problem complexity and the computational effort to solve the resulting optimization problem.
Therefore, we analyze the necessity of a thermal battery model as described in Sec.~\ref{ssec:thermal_model}.
In Mode III, assuming constant battery temperature, 
i.e.\ $\Delta \Theta_n = 0.0, \forall n \in \mathcal{N}$, 
would underestimate the operating cost by 0.55\% compared with a data-driven thermal model.
Applying Mode II, however, the operating cost would be underestimated by 3.44\%.

\begin{figure}[!t]
	\begin{footnotesize}
		\centering
		\tikzin{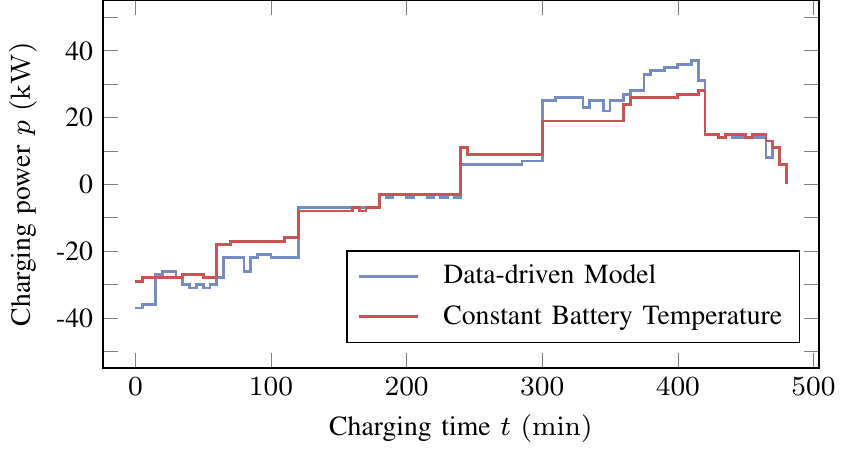}
		\caption{
			Charging power profiles over time for constant battery temperature (red) and
			data-driven thermal model (blue,  see also Sec.~\ref{ssec:thermal_model}).}
		\label{fig:temperature_influence}
	\end{footnotesize}
\end{figure}

Besides the errors in estimating the operating cost, the presence of a thermal model also influences the decision made by the optimization scheme, 
i.e. the charging power trajectory $\boldsymbol{p}^{*}$.
Figure \ref{fig:temperature_influence} shows exemplary power profiles over time assuming constant battery temperature and using a data-driven thermal model.
For $|p| > 7 \text{ kW}$, the mean difference of charging power is 3.11~kW, when comparing the constant battery temperature assumption with the data-driven thermal model. 
However, for $|p| \leq 7 \text{ kW}$ the mean deviation of charging power only amounts to 0.75~kW. 

Although the operating cost only show minor deviations, the charging power profiles change significantly.
In particular, the relevance of the battery temperature rises with the (absolute) charging power.
Hence, we suggest to use advanced thermal models, e.g. as in Sec.~\ref{ssec:thermal_model} for $|p| > 7 \text{ kW}$.
For $|p| \leq 7 \text{ kW}$, assuming constant battery temperature suffices.

\subsection{Effects of Battery Prices}
Due to EV market growth and battery technology improvements battery production prices will likely decrease within the next decade \cite{Sun2018, Berckmans2017, Hsieh2019}.%
\footnote{%
	Also advances in battery technology, i.e. reduced battery aging, would have similar effects as decreasing battery production prices.
}
In anticipation of SC for EV fleets, we compare EV operating cost of real charging events (see Sec.~\ref{sec:case_study}) with future battery prices for 2025 ($V_{\text{EV}} = \SI{4470}{\eur}$) and 2030 ($V_{\text{EV}} = \SI{2770}{\eur}$) taken from \cite{Sun2018}.%
\footnote{%
	To highlight the sole influence of decreasing battery prices, we assume electricity prices of 2018 as the cost calculation's underlying future electricity prices.
}
This directly affects \eqref{eq:aging_cost}, i.e. the aging costs $J_{\text{D}}$ could in average decrease by 26.5\% in 2025 and by 54.4\% in 2030 compared with 2020 battery prices ($V_{\text{EV}} = \SI{6080}{\eur}$).
Regarding the total operating cost, however, the decrease would only amount to 6.8\% in 2025, or 15.9\% in 2030, respectively.
This reduction of battery aging costs is not sufficient for \textit{energy arbitrage} to become profitable from a user's point of view.
A conceivable setup for power suppliers to incentivize EV owners to participate in V2G services could be a flat compensation for battery aging costs per charging event.

\subsection{Influence of Electricity Tariff}
Due to relatively small electricity price variations over time, see Fig.~\ref{fig:energy_price_profile}, many V2G services, e.g. \textit{energy arbitrage}, are unlikely to be profitable.
Depending on the charging event time window, the ratio
\begin{equation}
	\gamma = \frac{\epsilon_{\text{sell}}}{\epsilon_{\text{buy}}},
\end{equation} 
does not suffice to compensate for round-trip energy losses and aging costs.%
\footnote{%
	Note that $\epsilon_{\text{sell}}$ and $\epsilon_{\text{buy}}$ may not necessarily occur at the same time.
}
In a future use case, the grid operator may utilizes EV batteries as power reserve to compensate drastic power grid imbalances;
e.g., a charging event with electricity underproduction (high $\epsilon_{\text{sell}}$) at the beginning and electricity overproduction (low $\epsilon_{\text{buy}}$) towards the end. 
Conversion losses and battery aging costs must then be compensated accordingly.

\begin{figure}[!t]
	\begin{footnotesize}
		\centering	
		\tikzin{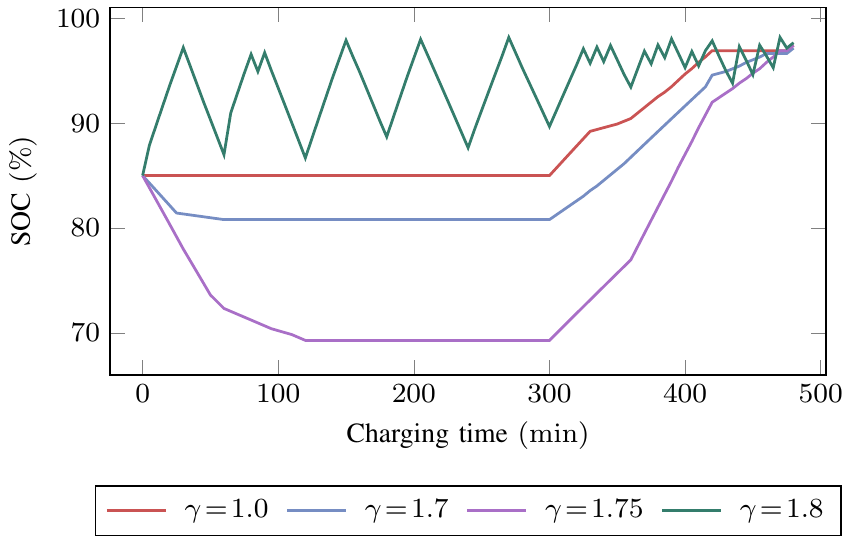}
		\caption{
			SOC profiles over time for different factors $\gamma$ of increased electricity selling price $\epsilon_{\text{sell}}$.
		}
		\label{fig:energy_profile_gamma}
	\end{footnotesize}
\end{figure}

\begin{figure}[!t]
	\begin{footnotesize}
		\centering	
		\tikzin{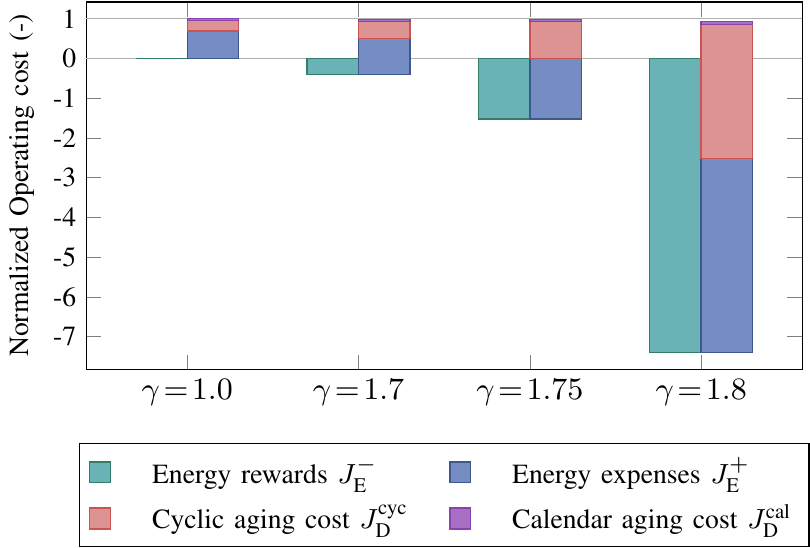}
		\caption{
			Normalized operating cost for different factors $\gamma$ of increased electricity selling price $\epsilon_{\text{sell}}$.
		}
		\label{fig:operating_cost_gamma}
	\end{footnotesize}
\end{figure}

To evaluate the impact on the optimization, we synthetically increase $\gamma$, i.e. simultaneously offer higher selling prices than buying prices. 
Figure \ref{fig:energy_profile_gamma} shows SOC profiles of four different $\gamma$.
Compared with the standard pricing $\gamma = 1.0$, the share of discharging phases, i.e. selling energy back to the grid, increases with $\gamma$.
Note the green SOC profile with $\gamma = 1.8$ cycling up and down various times, i.e. the battery is charged and discharged repeatedly. 
The optimization scheme thus fully utilizes the price differences for buying and selling energy.
Examining the operating cost and its components, as shown in Fig.~\ref{fig:operating_cost_gamma}, supports this result.
Energy rewards $J_{\text{E}}^{-}$, energy expenses $J_{\text{E}}^{+}$, and cyclic aging costs $J_{\text{D}}^{\text{cyc}}$ grow with $\gamma$.
Due to the $\gamma$-increased selling price $\epsilon_{\text{sell}}$, however, the energy rewards are growing more extensively than energy expenses and cyclic aging costs.
Thus, the total operating cost decline with increasing $\gamma$.
Again, note the drastic increase of $J_{\text{E}}^{+}$ when increasing $\gamma = 1.75$ to $\gamma = 1.8$.
Passing a specific value for $\gamma$, the energy rewards fully compensate round-trip energy losses and aging costs.
When aiming to spontaneously influence the EV charging process externally, e.g. as a power supplier, $\gamma$ must exceed this threshold.

To support future work on grid-supporting V2G services, we estimate a characteristic threshold for $\gamma$, see also \cite{Bishop2016, Carrion2019}.
Therefore, we assume charging the battery in one time interval and discharging in the second one with equal (absolute) power $|p| \leq 7\text{ kW}$.
The round-trip energy efficiency is represented by $\eta$.
We then obtain a characteristic threshold
\begin{equation}
	\gamma^* = \frac{J_\text{E} + 2 J_\text{D}}{\eta\cdot J_\text{E}}.
\end{equation}
With $\theta=21\text{ \si{\celsius}}$ and $\eta = 0.997$, we obtain $\gamma^* = 1.746$. 
Hence, for $\epsilon_{\text{sell}} > 1.746 \cdot \epsilon_{\text{buy}}$, the charging process can be influenced by price differences, given that the problem constraints still hold.
As we suppose electricity prices supplemented by fees and taxes (see Sec.~\ref{sec:case_study}), $\gamma$ is mostly less than $\gamma^*$.
For a more responsive SC control, an adapted price policy would require the fluctuations of the electricity market, e.g. extensive price peaks, to be passed to the EV customer. 

\section{Conclusion}\label{sec:conclusion}
In the present work we analyzed the influence of battery aging on Smart Charging (SC) of Electric Vehicles (EVs).
We modeled the EV battery using onboard sensor data and set up an optimization-based SC use case.
Evaluating the concept with real-world EV data revealed the need for advanced thermal models when charging power exceeds 7 kW.
We found that exploiting time and energy flexibility of EV charging reduces operating cost by 7.8\%.
Furthermore, disregarding battery aging underestimates EVs' operating cost up to 30\%.
Battery aging costs thus hinders many Vehicle-to-Grid (V2G) services based on bidirectional power flow from being profitable.
To overcome this would require a vast decrease of battery production prices or an adapted price policy with 75\% higher selling than electricity buying prices.
%
Future work will examine stochastic influences on SC.
These comprise EV user behavior, random V2G service allocation, integration of renewable energy sources, and dynamic constraints of transformers and charging stations.

\section*{Acknowledgment}
The authors would like to thank Mercedes-Benz AG (Sindelfingen, Germany) for the allocated resources and data required for this work.
SM, RM and VH are funded by the Helmholtz Association under the Program Energy System Design and the Joint Initiative ``Energy System 2050 – A Contribution of the Research Field Energy''.

\section*{Notation}\label{sec:notation}
\begin{itemize}
	\item 	$\hat{x}$ indicates an estimate of $x$
	\item 	$\underline{x}$ and $\overline{x}$ indicate the lower and upper bounds of $x$
	\item 	$x_n$ indicates the value of $x$ at time $t_n$
	\item 	$\boldsymbol{x}_{\text{d}}$ represents a discrete space between $\underline{x}$ and $\overline{x}$ for DDP\\
\end{itemize}
\subsection*{Acronyms}
\begin{tabularx}{\columnwidth}{l X}
	ANN &				Artificial Neural Network\\
	DDP &				Discrete Dynamic Programming\\
	ECM &				Equivalent Circuit Model\\
	EV & 				Electric Vehicle\\
	MAE & 				Mean Absolute Error\\
	RMSE & 				Root Mean Squared Error\\
	SC &				Smart Charging\\
	SOC &				State of Charge\\
	SOH & 				State of Health\\
	V2G &				Vehicle-to-Grid\\
	
\end{tabularx}
\subsection*{Parameters}
\begin{tabularx}{\columnwidth}{l X}
	$e_{\text{nom}}$ & 			Nominal available battery capacity (kWh)\\
	$H_{\text{EV}}$ & 	Total battery capacity fade, EV application (-)\\
	$\Delta t$ & Duration of discrete time interval (min)\\
	$V_{\text{EV}}$ & 	Total battery value loss, EV application (\euro)\\
	$\epsilon_{\text{buy}}$ & 	Electricity buying price (\euro)\\
	$\epsilon_{\text{sell}}$ & 	Electricity selling price (\euro)\\
	$\eta$ & 					Efficiency of charging process (-)\\
	$\lambda$ 							& DDP Penalty Cost (\euro)\\
	
\end{tabularx}
\subsection*{Variables}
\begin{tabularx}{\columnwidth}{l X}
	$e$ & 						Battery energy (kWh)\\
	$e_{\text{max}}$ & 			Maximum available battery capacity (kWh)\\
	$\Delta E$ & 				Energy throughput (kWh)\\
	$H$ &						Battery state of health (-)\\
	$\Delta H_{\text{cyc}}$ & 	Cyclic battery capacity fade (-)\\
	$\Delta H_{\text{cal}}$ & 	Calendar battery capacity fade (-)\\	
	$I_{\text{bat}}$ & 			Battery current (A)\\	
	$\boldsymbol{J}$			& DDP Cached Total Cost (\euro)\\
	$J_{\text{E}}$ & 			Energy cost function (\euro)\\
	$J_{\text{D}}$ & 			Battery degradation cost function (\euro)\\	
	$\boldsymbol{\mathfrak{J}}$		& DDP Cost Grid (\euro)\\
	$N$ & 						Number of time intervals (-)\\
	$\mathcal{N}$ & 			Set of time intervals (-)\\
	$p$ & 						Gross charging power (kW)\\
	$\boldsymbol{p}^{*}$ & 		Optimal charging power trajectory (kW)\\
	$\boldsymbol{\mathfrak{P}}$		& DDP Optimal Action Grid (kW)\\
	$\dot{Q}_\text{loss}$ & 	Heat flow from internal battery losses (kW)\\
	$R_\text{i}$ & 				Battery internal resistance (\si{\ohm})\\
	$U_\text{OCV}$ & 			Battery open-circuit voltage (V)\\
	$U_\text{bat}$ & 			Battery terminal voltage  (V)\\
	$\gamma$ & 					Electricity price selling-buying-ratio (-)\\
	$\theta$	&				Battery temperature (\si{\degreeCelsius})\\
	$\Delta \Theta$ & 			Battery temperature difference (K)\\	
	$\tau$ &  					Battery aging time equivalent (s)\\
\end{tabularx}
%
%
%

\appendix[Discrete Dynamic Programming Algorithms]
\begin{algorithm}
	\caption{%
		Initialization of backward induction algorithm (see Algorithm~\ref{alg:backward_ind_ddp}), acc.\ \cite{ArtLew2006}.
	}
	\label{alg:backward_ind_init_ddp}
	
	\Input{%
		$N,
		e_0,
		e_N,
		\theta_0,
		\underline{e},
		\overline{e},
		\underline{\theta},
		\overline{\theta},
		\underline{p},
		\overline{p},
		\lambda$
	}
	\Blank
	\Comment{discretize state and action:}
	\nl $\boldsymbol{e}_{\text{d}} \leftarrow	\mathtt{range}(\text{start: }\underline{e}, \text{stop: }\overline{e}, \text{step: } 0.8\text{ kWh})$\;
	\nl $\boldsymbol{\theta}_{\text{d}} \leftarrow	\mathtt{range}(\text{start: }\underline{\theta}, \text{stop: }\overline{\theta}, \text{step: }\SI{1}{\kelvin})$\;
	\nl $\boldsymbol{p}_{\text{d}} \leftarrow	\mathtt{range}(\text{start: }\underline{p}, \text{stop: }\overline{p}, \text{step: }\SI{1}{\kilo\watt})$\;
	\Comment{initialize cost grid and action grid (penalty value $\lambda = \SI{1000}{\eur}$):}
	\nl $\boldsymbol{\mathfrak{J}} 
	\leftarrow 
	\mathtt{zeros}(N, \mathtt{length}(\boldsymbol{e}_{\text{d}}), \mathtt{length}(\boldsymbol{\theta}_{\text{d}}))$\;
	\nl $\boldsymbol{\mathfrak{J}}[0,:,:], \boldsymbol{\mathfrak{J}}[N,:,:] 
	\leftarrow 
	\lambda, \lambda$\;
	\nl $\boldsymbol{\mathfrak{J}}[
	0,
	\mathtt{argmin}(|\boldsymbol{e}_{\text{d}} - e_0|),
	\mathtt{argmin}(|\boldsymbol{\theta}_{\text{d}} - \theta_0|)] 
	\leftarrow 0$\;
	\nl $\boldsymbol{\mathfrak{J}}[N,\mathtt{argmin}(|\boldsymbol{e}_{\text{d}} - e_N|), :] 
	\leftarrow 0$\;
	\nl $\boldsymbol{\mathfrak{P}} 
	\leftarrow 
	\mathtt{zeros}(N, \mathtt{length}(\boldsymbol{e}_{\text{d}}), \mathtt{length}(\boldsymbol{\theta}_{\text{d}}))$\;
	\Blank
	\Output{%
		$\boldsymbol{\mathfrak{J}},
		\boldsymbol{\mathfrak{P}},
		\boldsymbol{e}_{\text{d}},
		\boldsymbol{\theta}_{\text{d}},
		\boldsymbol{p}_\text{d}$
	}
\end{algorithm}
\begin{algorithm}
	\caption{%
		Backward induction algorithm to create cost grid $\boldsymbol{\mathfrak{J}}$ and corresponding optimal actions $\boldsymbol{\mathfrak{P}}$, acc.~\cite{ArtLew2006}.
	}
	\label{alg:backward_ind_ddp}
	\Input{%
		$N,
		\boldsymbol{\mathfrak{J}},
		\boldsymbol{\mathfrak{P}},
		\boldsymbol{e}_\text{d},
		\boldsymbol{\theta}_\text{d},
		\boldsymbol{p}_\text{d},
		\underline{e},
		\overline{e},
		\underline{\theta},
		\overline{\theta},
		\underline{p},
		\overline{p},
		\boldsymbol{\epsilon}_{\text{buy}},
		\boldsymbol{\epsilon}_{\text{sell}},
		\lambda$}
	\Blank
	\For{$n \leftarrow N\!-\!1$ \textbf{to} $0$}{%
		\ForAll{$e_i \in \boldsymbol{e}_{\text{d}}$}{%
			\ForAll{$\theta_j \in \boldsymbol{\theta}_{\text{d}}$}{%
				\Comment{initialize cached total cost:}
				\nl $\boldsymbol{J} \leftarrow \mathtt{ones}(\mathtt{length}(\boldsymbol{p}_{\text{d}})) \cdot \lambda$\;
				
				\ForAll{$p_k \in \boldsymbol{p}_{\text{d}}$}{%
					\Comment{validate charging power constraints \eqref{eq:decision_variable}:}
					\If{$\underline{p}(e_i, \theta_j) \leq p_k \leq \overline{p}(e_i, \theta_j)$}
					{
						\Comment{calculate state transitions (Sec.~\ref{ssec:electrical_model}, \ref{ssec:thermal_model}):}
						\nl $e_{n\!+\!1} \leftarrow e_i + \Delta E(e_i, \theta_j, p_k)$\;
						\nl $\theta_{n\!+\!1} \leftarrow \theta_j \! + \! \Delta \Theta(p_k, \dot{Q}_\text{loss}(e_i, \theta_j, p_k),$\;
						$\quad\Delta E(e_i, \theta_j, p_k), \theta_j)$\;
						\Comment{validate state constraints \eqref{eq:battery_energy} and \eqref{eq:battery_temperature}:}
						
						\If{$\underline{e} \leq e_{n\!+\!1} \leq \overline{e}$ \textbf{and} $\underline{\theta} \leq \theta_{n\!+\!1} \leq \overline{\theta}$}
						{%
							\Comment{calculate transition costs \eqref{eq:energy_cost} and \eqref{eq:aging_cost}:}
							\nl $J_\text{E} \leftarrow J_{\text{E}}(p_k, \epsilon_{\text{buy},n}, \epsilon_{\text{sell},n})$\;
							\nl $J_\text{D} \leftarrow \frac{V_{\text{EV}}}{H_{\text{EV}}}(\Delta H_{\text{cal}}(e_i, \theta_j, H_0)+$\;
							$\quad\Delta H_{\text{cyc}}(|\Delta E(e_i, \theta_j, p_k)|))$\;
							
							\Comment{calculate and cache total cost:}
							\nl $\boldsymbol{J}[k] \leftarrow J_\text{E} + J_\text{D} + \boldsymbol{\mathfrak{J}}[\mathtt{argmin}(|\boldsymbol{e}_{\text{d}} -\!$\;
							$\quad e_{n\!+\!1}|), \mathtt{argmin}(|\boldsymbol{\theta}_{\text{d}}\!-\!\theta_{n\!+\!1}|)]$\;	
						}
					}
				}
				\Comment{assign minimum cached cost and corresponding action:}
				\nl $\boldsymbol{\mathfrak{J}}[n,i,j] \leftarrow \mathtt{min}(\boldsymbol{J})$\;
				\nl $\boldsymbol{\mathfrak{P}}[n,i,j] \leftarrow \boldsymbol{p}_{\text{d}}[\mathtt{argmin}(\boldsymbol{J})]$\;
			}
		}
	}
	\Blank
	\Output{%
		$\boldsymbol{\mathfrak{J}},
		\boldsymbol{\mathfrak{P}}$
	}
\end{algorithm}
\begin{algorithm}
	\caption{%
		Forward integration algorithm to find the optimal charging power trajectory $\boldsymbol{p}^*$, acc.\ \cite{ArtLew2006}.
	}
	\label{alg:forward_int_ddp}
	\Input{%
		$N,
		\boldsymbol{\mathfrak{J}},
		\boldsymbol{\mathfrak{P}},
		e_0,
		\theta_0,
		\boldsymbol{e}_{\text{d}},
		\boldsymbol{\theta}_{\text{d}},
		\boldsymbol{\epsilon}_{\text{buy}},
		\boldsymbol{\epsilon}_{\text{sell}}
		$
	}
	\Blank
	\Comment{find starting point in the cost grid:}
	\nl $i, j \leftarrow \mathtt{argmin}(\boldsymbol{\mathfrak{J}}[0,:,;])$\;
	\Comment{initialize output and assign corresponding action:}
	\nl $\boldsymbol{p}^* \leftarrow \mathtt{zeros}(N)$\;
	\nl $\boldsymbol{p}^*[0] \leftarrow \boldsymbol{\mathfrak{P}}[0,i,j]$\;
	\Comment{initialize costs:}
	\nl $J_\text{E}, J_\text{D} \leftarrow 0,0$\;
	\Comment{start forward integration loop:}
	\For{$n \leftarrow 0$ \textbf{to} $N\!-\!1$}{%
		\Comment{calculate state transitions (see Sec.~\ref{ssec:electrical_model} and \ref{ssec:thermal_model}):}
		\nl $e_{n\!+\!1} \leftarrow e_n + \Delta E(e_n, \theta_n, \boldsymbol{p}^*[n])$\;
		\nl $\theta_{n\!+\!1} \leftarrow \theta_n + \Delta \Theta(\boldsymbol{p}^*[n], \dot{Q}_\text{loss}(e_n, \theta_n, \boldsymbol{p}^*[n]),$\;
		$\quad\Delta E(e_n, \theta_n, \boldsymbol{p}^*[n]), \theta_n)$\;
		\Comment{calculate costs with \eqref{eq:energy_cost} and \eqref{eq:aging_cost}:}
		\nl $J_\text{E} \leftarrow J_{\text{E}} +J_{\text{E}}(p_n, \epsilon_{\text{buy},n}, \epsilon_{\text{sell},n})$\;
		\nl $J_\text{D} \leftarrow J_\text{D} + \frac{V_{\text{EV}}}{H_{\text{EV}}}( \Delta H_{\text{cal}}(e_n,\theta_n,H_0))+$\;
		$\quad \Delta H_{\text{cyc}}(|\Delta E(e_n, \theta_n, \boldsymbol{p}^*[n])|) $\;
		\Comment{find nearest discrete state and assign corresponding action:}
		\nl $\boldsymbol{p}^*[n\!+\!1] \leftarrow \boldsymbol{\mathfrak{P}}[n\!+\!1, \mathtt{argmin}(|\boldsymbol{e}_{\text{d}}\!-\!e_{n\!+\!1}|),$\; $\quad\mathtt{argmin}(|\boldsymbol{\theta}_{\text{d}}\!-\!\theta_{n\!+\!1}|)]$\;
	}
	\Output{%
		$\boldsymbol{p}^*,
		J_\text{E}, 
		J_\text{D}
		$}
\end{algorithm}
%
%
%
%
%
%
\newpage

%
%
%
%
%
%

%
%
%
%
%
%

%
%
%
\ifCLASSOPTIONpeerreview
	
\else

\begin{IEEEbiography}[{\includegraphics[width=1in,height=1.25in,clip,keepaspectratio]{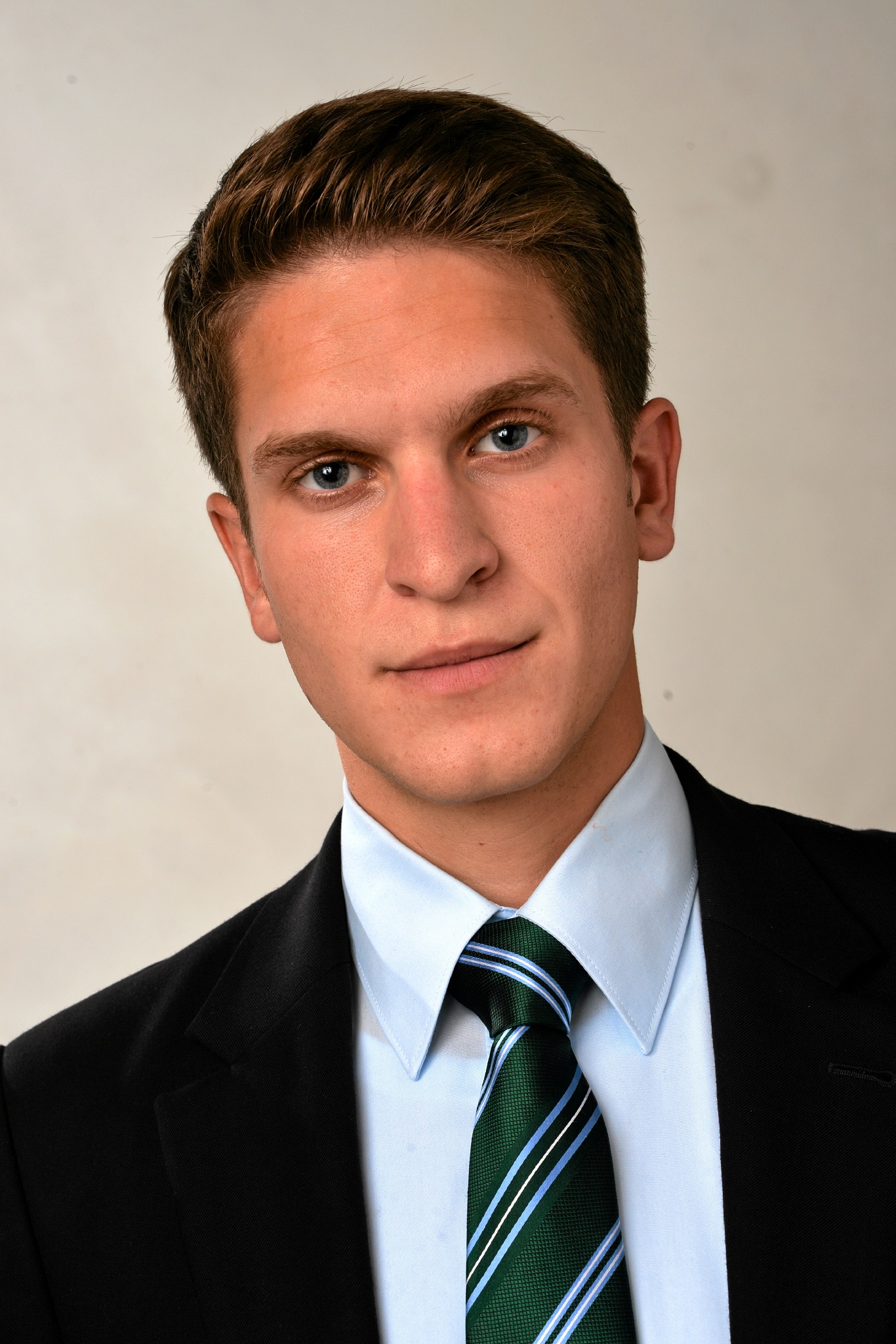}}]{Karl Schwenk}
	received the M.Sc. degree in mechanical engineering from the Karlsruhe Institute of Technology, Karlsruhe, Germany in 2018.
	He is currently pursuing his Ph.D. at the Institute for Automation and Applied Informatics with the Karlsruhe Institute of Technology in cooperation with Mercedes-Benz AG, Sindelfingen, Germany.
	His research topics comprise automated charging assistants for electric vehicles considering power supply, battery degradation and user constraints.
\end{IEEEbiography}
\begin{IEEEbiography}[{\includegraphics[width=1in,height=1.25in,clip,keepaspectratio]{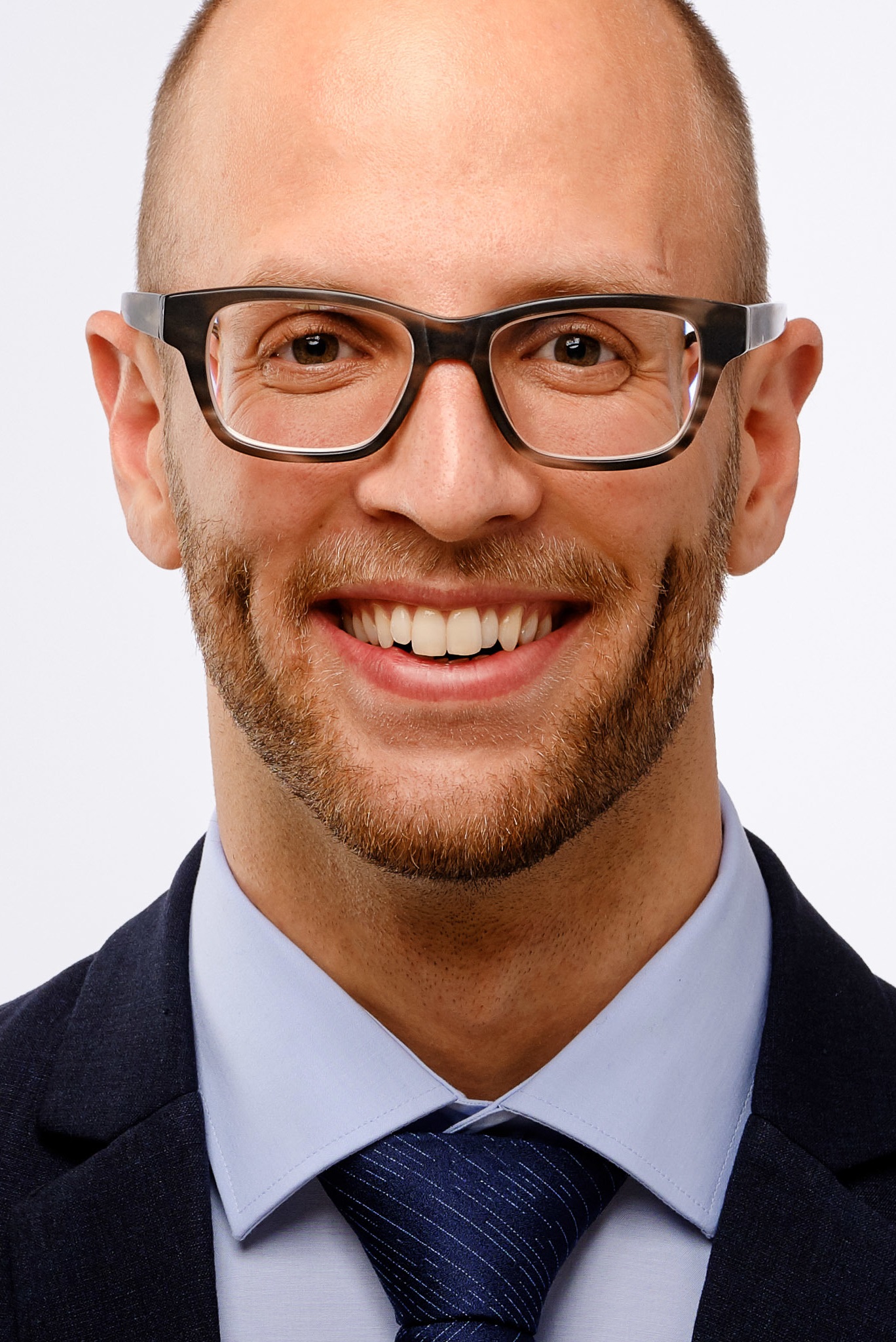}}]{Stefan Meisenbacher}
	received the B.Eng. in mechanical engineering from the Esslingen University of Applied Sciences, Esslingen, Germany in 2017, and the M.Sc. degree in mechanical engineering from the Karlsruhe Institute of Technology, Karlsruhe, Germany in 2020. 
	He is currently pursuing a Ph.D. in the field of modeling, machine-learning based forecasting and model application surveillance at the Institute for Automation and Applied Informatics within the Karlsruhe Institute of Technology.
\end{IEEEbiography}
\begin{IEEEbiography}[{\includegraphics[width=1in,height=1.25in,clip,keepaspectratio]{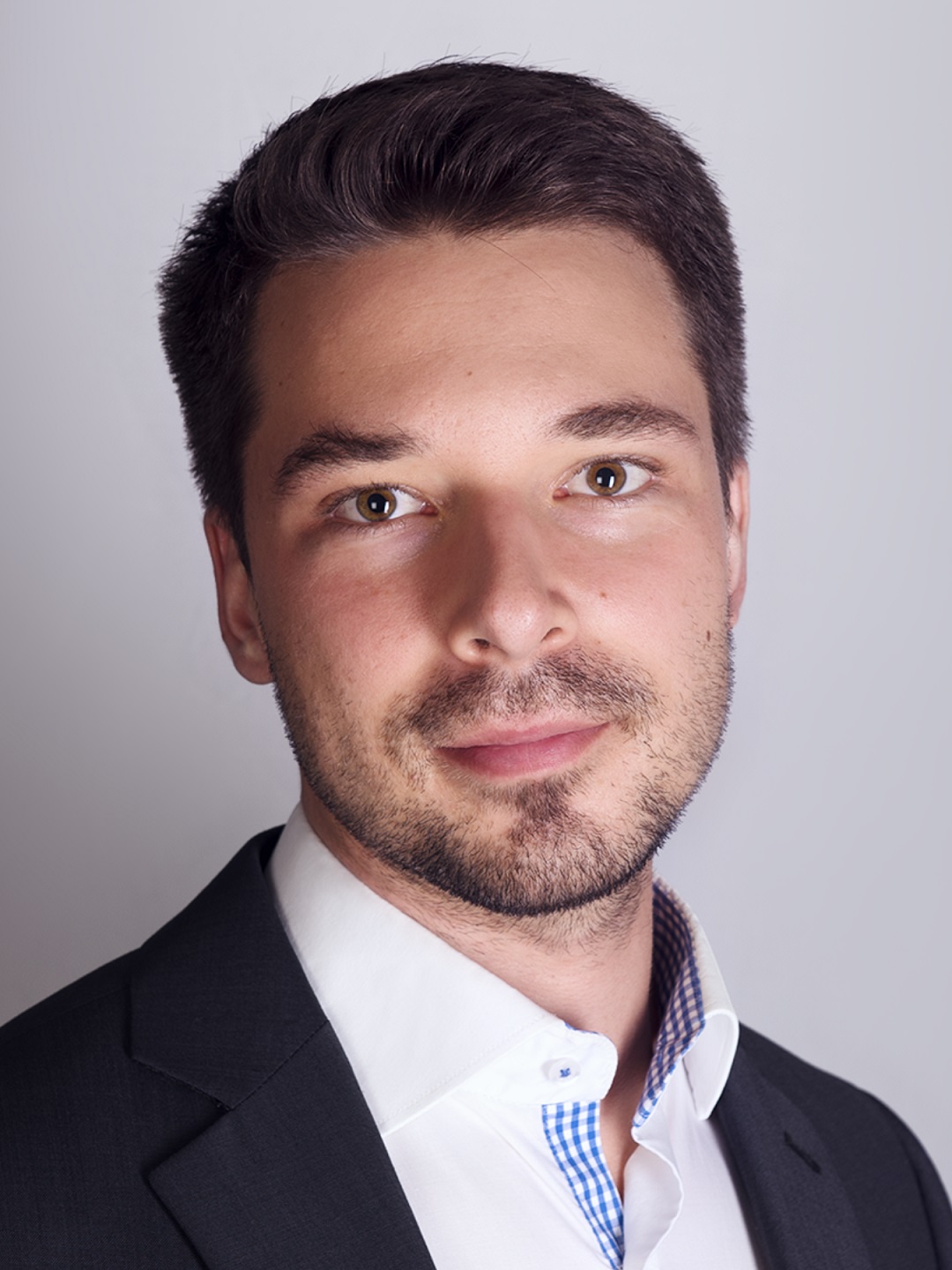}}]{Benjamin Briegel}
	received the Dipl.-Ing. degree in mechanical engineering from the University of Stuttgart, Stuttgart, Germany, in 2014. 
	He is currently working as a development engineer in the area of eDrive innovations at Mercedes-Benz AG in Sindelfingen, Germany. 
	His work topics include the development of digital services for electric vehicles with a focus on optimal charging.
\end{IEEEbiography}
\begin{IEEEbiography}[{\includegraphics[width=1in,height=1.25in,clip,keepaspectratio]{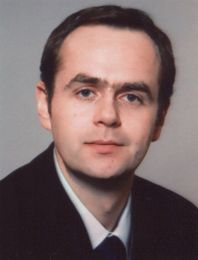}}]{Tim Harr}
	received the Ph.D. degree in electrical engineering from the University of Ulm, Ulm, Germany, in 2007. 
	He is currently working as a development engineer in the area of eDrive innovations at Mercedes-Benz AG in Sindelfingen, Germany. 
	His work topics include the development of connected services to detect and predict the usage behavior of electric vehicles.
\end{IEEEbiography}
\begin{IEEEbiography}[{\includegraphics[width=1in,height=1.25in,clip,keepaspectratio]{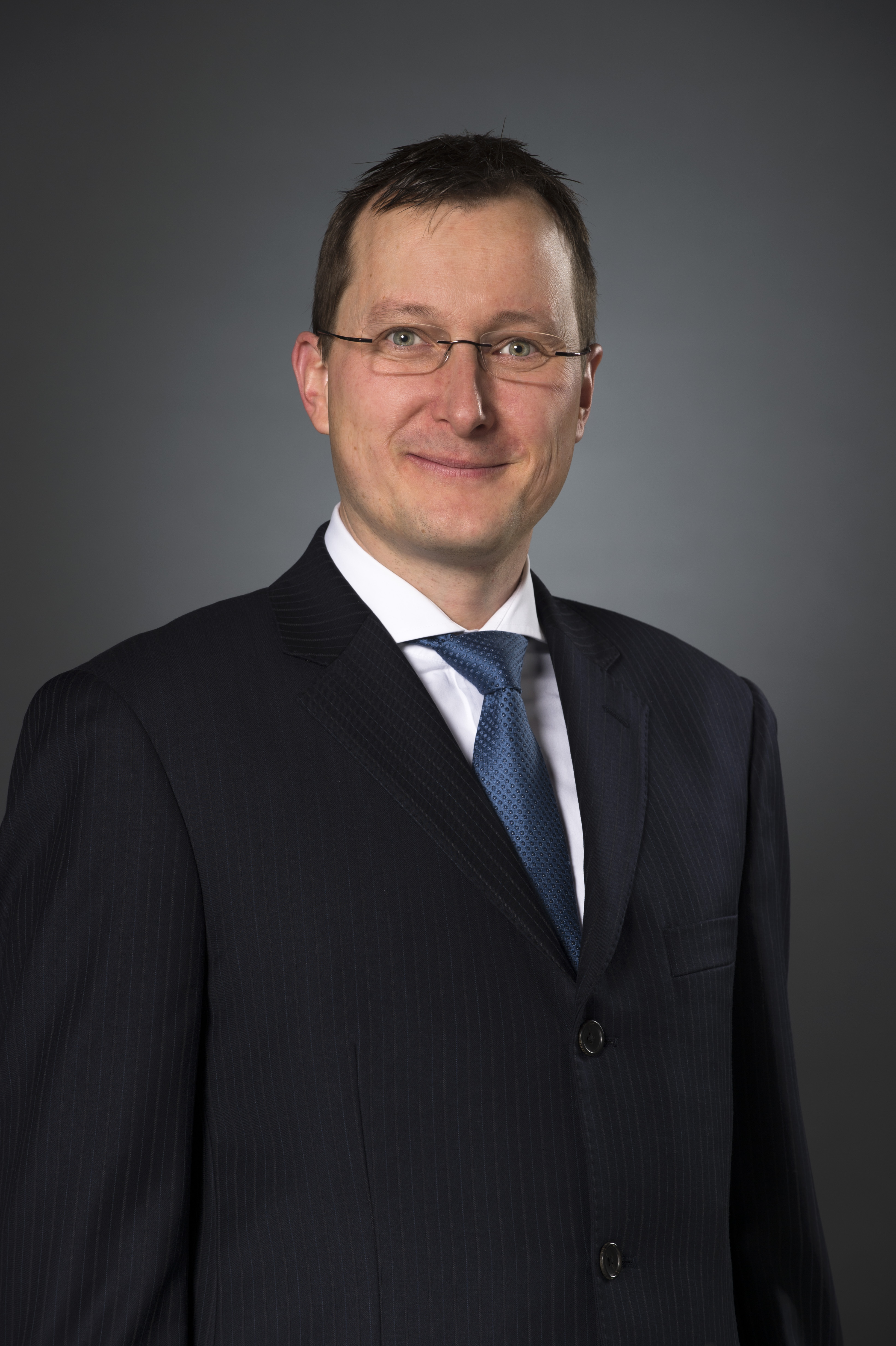}}]{Veit Hagenmeyer}
	received the Ph.D. degree from Université Paris XI, Paris, France in 2002. 
	He is currently the Professor in energy informatics with the Faculty of Informatics, and the Director of the Institute for Automation and Applied Informatics with the Karlsruhe Institute of 			Technology, Karlsruhe, Germany. 
	His research interests include modeling, optimization and control of sector-integrated energy systems, machine-learning based forecasting of uncertain demand and production in energy systems 			mainly driven by renewables, and integrated cyber-security of such systems.
\end{IEEEbiography}
\begin{IEEEbiography}[{\includegraphics[width=1in,height=1.25in,clip,keepaspectratio]{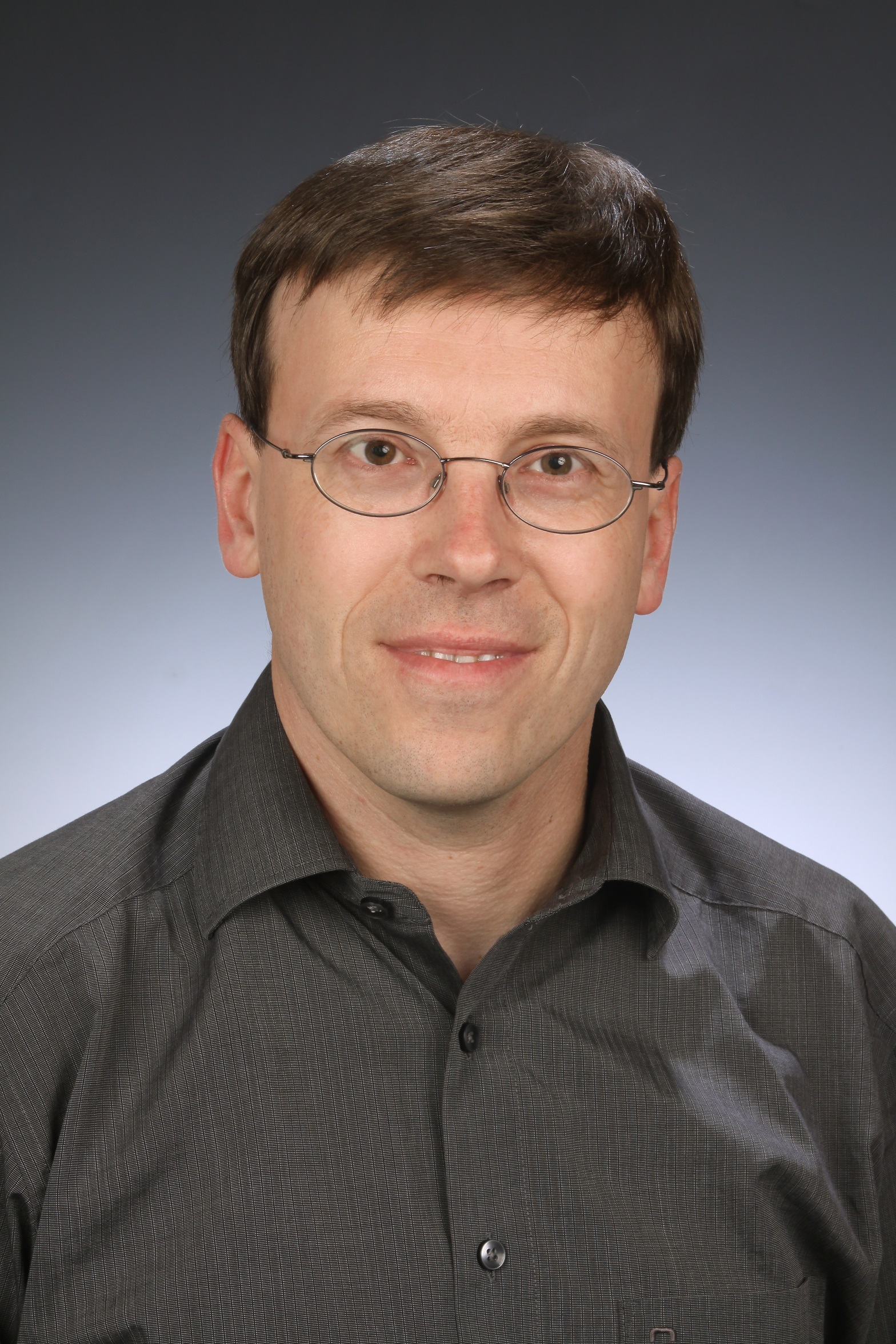}}]{Ralf Mikut}
	received the Dipl.-Ing. degree in automatic control from the University of Technology, Dresden, Germany, in 1994, and the Ph.D. degree in mechanical engineering from the University of Karlsruhe, 	Karlsruhe, Germany, in 1999.
	Since 2011, he is Adjunct Professor at the Faculty of Mechanical Engineering and Head of the Research Group “Automated Image and Data Analysis” at the Institute for Automation and Applied 			Informatics of the Karlsruhe Institute of Technology (KIT), Germany. 
	His current research interests include machine learning, image processing, life science applications and smart grids.
\end{IEEEbiography}
%
%
%

\fi
\end{document}